\documentclass[12pt]{article}
\usepackage{graphicx}
\usepackage{array}
\usepackage{multirow}
\usepackage{amsfonts}
\usepackage{ulem}
\usepackage{color}
\usepackage{amssymb}
\usepackage{mathrsfs}
\usepackage{amsmath}
\usepackage{subfigure}
\makeatletter
\newcommand{\rmnum}[1]{\romannumeral #1}
\newcommand{\Rmnum}[1]{\expandafter\@slowromancap\romannumeral #1@}
\makeatother

\usepackage[unicode]{hyperref}
\usepackage[numbers,sort&compress]{natbib}

\begin{document}
\renewcommand{\thefootnote}{\fnsymbol{footnote}}
\begin{titlepage}

\vspace{10mm}
\begin{center}
{\Large\bf Higher-dimensional regular Reissner-Nordstr\"{o}m black holes associated with linear electrodynamics}
\vspace{9mm}

{{\large Yu-Mei Wu${}^{1,2,}$\footnote{E-mail address: wuym@mail.nankai.edu.cn} and Yan-Gang Miao${}^{1,3,}$\footnote{Corresponding author. E-mail address: miaoyg@nankai.edu.cn}}\\

\vspace{6mm}
${}^{1}${\normalsize \em School of Physics, Nankai University, Tianjin 300071, China}

\vspace{3mm}
${}^{2}${\normalsize \em College of Arts and Sciences, Fuzhou Institute of Technology, Fuzhou 350506, China}

\vspace{3mm}
${}^{3}${\normalsize \em Bethe Center for Theoretical Physics and Institute of Physics, University of Bonn, \\
Nussallee 12, 53115 Bonn, Germany}
}

\end{center}

\vspace{10mm}
\centerline{{\bf{Abstract}}}
\vspace{6mm}
\noindent

 Following the interpretation of matter source that the energy-momentum tensor of anisotropic fluid can be dealt with effectively as the energy-momentum tensor of perfect fluid plus linear (Maxwell) electromagnetic field,
we obtain the regular higher-dimensional Reissner-Nordstr\"{o}m (Tangherlini-RN) solution
by starting with the noncommutative geometry inspired Schwarzschild solution.
Using the boundary conditions that connect the noncommutative Schwarzschild solution in the interior of the  charged perfect fluid sphere to the Tangherlini-RN solution in the exterior of the sphere, we find that the interior structure can be reflected by the exterior parameter, the charge-to-mass ratio. 
Moreover, we investigate the stability of the boundary under mass perturbation and indicate that the new interpretation imposes a rigid restriction upon the charge-to-mass ratio. This restriction, in turn, permits  a stable noncommutative black hole only in the 4-dimensional spacetime.

\vskip 20pt
\noindent
{\bf PACS Number(s)}: 04.20.Cv, 04.40.Nr, 04.50.Gh


\end{titlepage}
\newpage
\renewcommand{\thefootnote}{\arabic{footnote}}
\setcounter{footnote}{0}
\setcounter{page}{2}

\section{Introduction}

The singularity puzzle appeared just after the general relativity was born, where the singularities of spacetime are most bewildering at the end of stars' collapsing and at the beginning of universe. It was believed that the singularities were the result of high symmetry, for instance, the collapsing stars' singularity originates from the spherical symmetry. On the other hand, the Hawking-Penrose singularity theorem proved~\cite{singularity,theorem} that the spacetime where gravity is strong enough would result in singularities under the causality condition and energy condition regardless of the local non-uniformity. Though the theorem is fabulous and convincing as well, it might not be the final answer to the puzzle. First of all, quantum theories are indispensable for space at microscales where the singularity lies, and a complete quantum theory of gravity is still lacking which 
may depict a different picture. Second of all, though the causality condition must be obeyed, the ways to avoid singularities have been found in different energy conditions.

The singularity could be replaced~\cite{sakharov,gliner} by a de Sitter core when the matter with high density would make a phase transition into some false vacuum state in the late time of collapsing. One further advance was the first construction of a nonsingular black hole (regular black hole or black hole with singularities free) by Bardeen~\cite{bardeen}. This model is asymptotically approaching a de Sitter phase at the center and asymptotically flat at infinity, and its energy-momentum tensor satisfies the weak energy condition (WEC) rather than the strong energy condition (SEC). Since then a great deal of research has been carried out on this issue, among which one of the valuable attempts is the interpretation of matter source. That is to say, the source originates~\cite{BNEI} from either anisotropic fluids or nonlinear magnetic monopoles in the Bardeen's model. For those models~\cite{dymnikova,hayward,1609.01629} based on anisotropic fluids,
the noncommutative geometry inspired Schwarzschild black hole~\cite{0510112,0807.1939} is of particular interest, where the anisotropic fluid appears naturally when the Gaussian distribution is taken as the black hole mass density due to the noncommutativity of spacetime. In this way, its metric resembles that of the Schwarzschild black hole but has a regular center. A nonsingular charged black hole solution was also found when both the mass and the charge take Gaussian distribution~\cite{0801.3519}.
As to nonsingularity realized by coupling gravitational field with nonlinear electrodynamic field, the solutions have been found~\cite{1408.0306, 1709.09473} satisfying the WEC and behaving asymptotically as the Reissner-Nordstr\"{o}m (RN) black hole. In addition, the other regular black hole solutions have also been obtained~\cite{Frolov,Balbinot,Barrabes} in terms of boundary soldering of two spacetimes. Along this way, the de Sitter spacetime was connected~\cite{1104.4790} with the RN spacetime through a charged spherical shell inside the RN inner horizon, and this junction technique was further developed~\cite{1209.3567} to form a massive thin layer.

In recent years, a new interpretation of matter source was suggested~\cite{1706.03454} for constructing the regular RN black hole solution, where the linear electrodynamics instead of nonlinear one was adopted.
To be specific, the regular solution can be interpreted as an exact solution of the Einstein equation coupled to linear electrodynamics if the components of the energy-momentum tensor satisfy the dominant energy condition (DEC). In other words, the energy-momentum tensor of anisotropic fluids can be dealt with effectively as the energy-momentum tensor of the perfect fluid plus linear electromagnetic field. As an application, the
nonsingular RN black hole solution was constructed in the 4-dimensional spacetime~\cite{1706.03454}, where the singularity at the RN center was replaced by a charged perfect fluid sphere\footnote{As was denoted in ref.~\cite{1706.03454}, the charged perfect fluid sphere, or a charged perfect fluid in short, means the perfect fluid together with the electromagnetic field.} located inside the RN inner horizon when a fixed range of charge-to-mass ratio was taken.

Moreover, the higher-dimensional spacetimes have particular interest to physicists due to some specific reasons:
\begin{itemize}
\item String theory~\cite{polkin}, the candidate to unify all fundamental interactions, is established in higher-dimensional spacetimes.
\item The AdS/CFT correspondence~\cite{9905111} relates quantum gravity theories in the $n$-dimensional spacetime to gauge field theories in the $(n-1)$-dimensional spacetime, which is the most successful realization of the holographic principle.
\item Large extra dimensions might provide~\cite{0210296} a possibility to detect non-perturbative gravitational objects such as black holes and branes in TeV scale in the future colliders.
\end{itemize}

The combination of the new interpretation~\cite{1706.03454} and higher dimensions leads to our motivation, i.e., our idea is to extend this interpretation related to linear electrodynamics to higher-dimensional spacetimes. 
It is of interest to study whether higher dimensions impose specific constraints upon the charge-to-mass ratio for the formation of a black hole and to analyze the stability of the formed black hole from the hole's shell that connects two spacetimes. 
The model we begin with is the noncommutative geometry inspired Schwarzschild black hole with the Gaussian source~\cite{0811.2685} in higher-dimensional spacetimes. On one hand,  the noncommutative black holes are a remaining candidate to be expected to test~\cite{1801.05023} TeV-scale gravity at the LHC; and on the other hand, the well-motivated model and its extensions have been widely investigated from the interpretation of anisotropic fluid ~\cite{0606051, 0801.3519,1511.00853} and can thus provide us direct comparisons between different interpretations.

The present paper is organized as follows. In section 2, we decompose the energy-momentum tensor of a general higher-dimensional metric of anisotropic fluids into the energy-momentum tensor of the perfect fluid plus linear electromagnetic field, and derive the boundary conditions that relate the parameters of exterior solutions with the interior source distribution.  In section 3, we construct the regular Tangherlini-RN solution in the exterior of the charged perfect fluid sphere
by choosing the noncommutative geometry inspired Schwarzschild solution in the interior of the sphere, find out constraints on the charge-to-mass ratio under the hoop conjecture, and then discuss the stability of the boundary. Finally, we give a brief summary in section 4.

We adopt the following units throughout the paper: $c=1$ and $\mu_0=1/\epsilon_0=4\pi$, where $c$, $\mu_0$ and $\epsilon_0$ are the speed of light, the permeability of vacuum and the permittivity of vacuum, respectively.

\section{General formulation}
\subsection{General higher-dimensional nonsingular metrics}
A general static and spherically symmetric metric in $n$ dimensions $(n\geq4)$ reads
\begin{align}
\mathrm{d}s^2=-f(r)\mathrm{d}t^2+\frac{1}{f(r)}\mathrm{d}r^2+r^2\,\mathrm{d}\Omega_{n-2}^2,
\label{metric}
\end{align}
where $\mathrm{d}\Omega_{n-2}$ is the line element of an $(n-2)$-dimensional unit sphere.  In General Relativity, the singularity under the metric often manifests as curvature infiniteness at $r=0$. Accordingly, being free of curvature singularities requires that the invariants associated with the Riemann curvature tensor are regular everywhere. Under this metric, these invariants can be expressed as follows,
\begin{align}
&R=-f''(r)+\frac{(n-2)(n-3)}{r^2}\left[1-f(r)\right], \notag \\
&R_{\mu\nu}R^{\mu\nu}=\frac{1}{2}\left[f''(r)+\frac{(n-2)f'(r)}{r}\right]^2+\frac{n-2}{r^4}\left[(n-3)(1-f(r))-rf'(r)\right]^2, \notag \\
&R_{\mu\nu\gamma\delta}R^{\mu\nu\gamma\delta}=\left[f''(r)\right]^2+2(n-2)\frac{\left[f'(r)\right]^2}{r^2}+\frac{2(n-2)(n-3)}{r^4}\left[1-f(r)\right]^2.\label{scalars}
\end{align}

Moreover, the energy-momentum tensor that depicts the matter source with the spherical symmetry can be written as
\begin{equation}
{T^\mu}_\mu={\rm diag}(-\rho, p_r, \underbrace{p_\bot, p_\bot, \cdots, p_\bot}_{n-2}),
\label{EMT}
\end{equation}
where $\rho$ is the density, $p_r$ the radial pressure, and $p_\bot$ the angular pressure. Such a source is usually regarded as an anisotropic fluid.
By solving the Einstein equation,
\begin{align}
G_{\mu \nu}=R_{\mu \nu}-\frac{1}{2} g_{\mu \nu}R=\kappa_n T_{\mu \nu},
\end{align}
where $\kappa_n$ is Einstein's constant in the $n$-dimensional spacetime, we obtain the components of the energy-momentum tensor,
\begin{align}
\rho&=\frac{n-2}{2\kappa_n r^2}\left\{(n-3)[1-f(r))]-r f'(r)\right\}, \label{rho} \\
p_r&=-\frac{n-2}{2\kappa_n r^2}\left\{(n-3)[1-f(r)]-r f'(r)\right\}, \label{p_r} \\
p_\bot&=\frac{1}{2\kappa_n r^2}\left\{r^2f''(r)-(n-3)\left[(n-4)(1-f(r))-2rf'(r)\right]\right\},
\label{p_bot}
\end{align}
and further find the relations between the components,
\begin{align}
\rho&=-p_r, \label{rhop} \\
p_\bot&=-\rho -\frac{r\rho'}{n-2}=p_r +\frac{r p_r'}{n-2},\label{pbot}
\end{align}
where a prime denotes the derivative with respect to $r$. We summarize the following three properties that the general higher-dimensional nonsingular black hole has:

 $\mathrm{(\rmnum{1})}$  The curvature invariants are regular everywhere due to the regularity of the metric, see eqs.~(\ref{metric}) and (\ref{scalars}), from which we can deduce that the functions, $(1-f(r))/r^2$, $f'(r)/r$ and $f''(r)$, behave regularly everywhere and that leads to the regular components of the energy-momentum tensor in eqs.~(\ref{rho}), (\ref{p_r}), and (\ref{p_bot}); 
 
 $\mathrm{(\rmnum{2})}$ $p_r\neq p_\bot$ (except at $r=0$), which gives~\cite{1501.07044} the reason for the ``anisotropic fluid" interpretation, and $\rho=-p_r\simeq -p_\bot$ near the center, indicating that the source is exactly in a de Sitter vacuum state; 
 
 $\mathrm{(\rmnum{3})}$ $\rho>0$ and $\rho'\leq 0$, as most of the regular black hole models meet, lead to the satisfaction of WEC but usually local violation of SEC.\footnote{For the energy-momentum tensor in eq.~(\ref{EMT}), the weak energy condition (WEC) requires $\rho>0$, $\rho+p_r\geq0$, and $\rho+p_\bot\geq0$; the strong energy condition (SEC) requires $\rho+p_r+(n-2)p_\bot\geq0$, $\rho+p_r\geq0$, and $\rho+p_\bot>0$.}

\subsection{New interpretation of matter source associated with linear electrodynamics}
The energy-momentum tensor (eq.~(\ref{EMT})) that is compatible with the nonsingular metric (eqs.~(\ref{metric}) and (\ref{scalars})) in a charged black hole model was usually interpreted~\cite{1408.0306, 1709.09473} by nonlinear electromagnetic field. Nonetheless, it can also be obtained~\cite{1706.03454} within the linear electrodynamic framework in the 4-dimensional spacetime based on the multicomponent fluid combination~\cite{1501.07044}. The key point is the decomposition of eq.~(\ref{EMT}) into the energy-momentum tensor of the perfect fluid plus Maxwell electromagnetic field. Under such a decomposition, the boundary of source forms, on which the density of the charged perfect fluid sphere vanishes, and thus it naturally connects the interior spacetime to the exterior one.
Next we follow the procedure suggested in ref.~\cite{1706.03454} and develop it to the higher-dimensional case.

We begin with decomposing the energy-momentum tensor  (eq.~(\ref{EMT})) into two parts,
\begin{align}
T_{\mu\nu}=\tau_{\mu\nu}+\varepsilon_{\mu\nu},
\label{EMT2}
\end{align}
where $\tau_{\mu\nu}$ is the energy-momentum tensor of the perfect fluid and $\varepsilon_{\mu\nu}$ the energy-momentum tensor of  the electromagnetic tensor.
The two tensors can be written as follows under our specific units mentioned at the end of section 1,
\begin{align}
\tau_{\mu\nu}&=(\rho_m+p)u_\mu u_\nu+p g_{\mu\nu},\\
\varepsilon_{\mu\nu}&=\frac{1}{4\pi}\left(F_{\mu\lambda}F_{\nu\sigma}g^{\lambda\sigma}-\frac{1}{4}g_{\mu\nu}F_{\lambda\sigma}F^{\lambda\sigma}\right),
\end{align}
where $u_\mu$, $\rho_m$ and $p$ represent the velocity, density and isotropic pressure of the perfect fluid, respectively,
and $F_{\mu\nu}$ is the electromagnetic field strength defined via the electromagnetic potential $A_\mu$,
\begin{align}
F_{\mu\nu}=\nabla_\mu A_\nu-\nabla_\nu A_\mu.
\end{align}
Moreover, the field strength satisfies the Maxwell equation,
\begin{align}
\nabla^\mu F_{\mu\nu}=-4\pi J_\nu,
\label{ME}
\end{align}
where $J_\mu$ is the current density defined with the electric charge density $\rho_e$ as follows,
\begin{align}
J_\mu :=\rho_e u_\mu.
\end{align}

For a comoving observer, the velocity takes a simpler form,
\begin{align}
u_\mu=-\sqrt{f(r)}\delta^t_\mu,
\end{align}
where $\delta$ is the Kronecker symbol, and $A_\mu$ takes the following form,
\begin{align}
A=A_\mu \mathrm{d}x^\mu=-\phi(r)\mathrm{d}t.
\end{align}

Namely, there exists only the electrostatic field $\phi(r)$ but no magnetic field. As a result, we obtain the nonvanishing components of $F_{\mu\nu}$ and $\varepsilon_{\mu\nu}$ as follows,
\begin{align}
F_{10}&=-F_{01}=-\phi'(r)=E,\\
\varepsilon_{00}&=\frac{E^2f(r)}{8\pi},\qquad \varepsilon_{11}=-\frac{E^2}{8\pi f(r)}, \qquad \varepsilon_{ii}=\frac{E^2r^2}{8\pi}, \qquad i=2,3,\cdots, n,
\end{align}
and further reduce the energy-momentum tensor of eq.~(\ref{EMT2}) to a concise form,
\begin{align}
{T^\mu}_\mu={\rm diag}(-\rho_m-W, p-W, \underbrace{p+W, p+W, \cdots, p+W}_{n-2}),
\label{EMT3}
\end{align}
where $W=E^2/(8\pi)$ is the energy density of the electrostatic field. This quantity is also called the effective energy-momentum tensor, i.e., a kind of effectively treated formulations from eq.~(\ref{EMT}).

Now the energy-momentum tensor presents two different formulations, one from the viewpoint of anisotropic fluid, see eq.~(\ref{EMT}), and the other from the viewpoint of perfect fluid plus Maxwell electromagnetic field, see eq.~(\ref{EMT3}). By combining the two formulations, we establish the relation between the two different points of view, that is, the components of the energy-momentum tensor based on the latter viewpoint can be written in terms of the components of the energy-momentum tensor based on the former one,
\begin{align}
& \rho_m=-p=\frac{1}{2}(\rho-p_\bot),\label{rhom}\\
& W=\frac{1}{2}(\rho+p_\bot).
\label{BC}
\end{align}

Because the perfect fluid and the electromagnetic field satisfy the weak energy condition separately, we deduce $\rho-p_\bot\geq0$ and $\rho+p_\bot\geq0$. That is, the energy-momentum tensor (eq.~(\ref{EMT})) naturally meets the dominant energy condition (DEC).\footnote{For the energy-momentum tensor eq.~(\ref{EMT}), the dominant energy condition requires $\rho\geq|p_r|$ and $\rho\geq |p_\bot|$, where the former inequality is satisfied due to eq.~(\ref{rhop}), and the latter due to the WEC.} It is notable that the radius at $\rho=p_\bot$ marks the boundary of the charged  perfect fluid sphere, since the density and pressure of the perfect fluid vanish there, cf. eq.~(\ref{rhom}).
Therefore, under the new interpretation of matter source that was proposed in four dimensions~\cite{1706.03454},  we can obtain the general nonsingular metric eq.~(\ref{metric}) by solving the Einstein equation coupled to linear electrodynamics in $n$ dimensions, and then express the mass density, the isotropic pressure, and the energy density of the charged perfect fluid in terms of the metric function $f(r)$ as follows,
\begin{align}
& \rho_m=-p=\frac{1}{4\kappa_n r^2}\left[2(n-3)^2-2(n-3)^2f(r)-(3n-8)r f'(r)-r^2 f''(r)\right],
\label{rho_m}\\
& W=\frac{1}{4\kappa_n r^2}\left[2(n-3)-2(n-3)f(r)+(n-4)r f'(r)+r^2 f''(r)\right].
\label{EME}
\end{align}

Further, we give the charge through solving eq.~(\ref{ME}),
\begin{align}
q(r)\equiv A_{n-2}\int_0^r \frac{\rho_e}{\sqrt{f(r)}}r^{n-2}\mathrm{d}r=A_{n-2}\frac{Er^{n-2}}{4\pi},
\label{charge}
\end{align}
where $A_{n-2}$ stands for the volume of an $(n-2)$-dimensional unit sphere. In addition, we reduce the conservation equation $\nabla_\nu T^{\mu\nu}=0$ to
\begin{align}
2p'+\frac{f'(r)}{f(r)}(\rho_m+p)=2\frac{E \rho_e}{\sqrt{f(r)}},
\end{align}
or more explicitly to $p'=\frac{E \rho_e}{\sqrt{f(r)}}$ by considering $\rho_m+p=0$, manifesting that the gradient of pressure is balanced by the electrostatic force as pointed out in the 4-dimensional spacetime~\cite{1706.03454}.

\subsection{Boundary conditions}
As mentioned above, we have $\rho_m(r_s)=-p(r_s)=0$ at the boundary $r=r_s$, where $r_s$ can be solved from eq.~(\ref{rho_m}) for a specific distribution $f(r)$. The boundary connects the internal spacetime of the charged perfect fluid sphere to the external 
one, where the internal solution is the Schwarzschild's with the Gaussian distribution of matter source and the external solution is the Tangherlini-RN form.
For the convenience in the next section, we give the boundary conditions.

The Tangherlini-RN metric reads~\cite{Tangherlini}
\begin{align}
\mathrm{d}s^2=-f_{RN}(r)\mathrm{d}t^2+\frac{1}{f_{RN}(r)}\mathrm{d}r^2+r^2\,\mathrm{d}\Omega_{n-2}^2,
\end{align}
with
\begin{align}
f_{RN}(r)=1-\frac{2\kappa_n}{(n-2)A_{n-2}}\frac{M}{r^{n-3}}+\frac{4\pi \kappa_n}{(n-2)(n-3)A_{n-2}^2}\frac{Q^2}{r^{2(n-3)}},
\label{f_rn}
\end{align}
where $M$ and $Q$, the total mass and charge, respectively, are determined by the matter inside the charged perfect fluid sphere. Specifically, two connecting conditions are required at the boundary $r_s$ in order to fix $M$ and $Q$, that is, both the metric function and the electric field need to be continuous at $r_s$,
\begin{align}
f(r_s)=f_{RN}(r_s)=f_s, \qquad E(r_s)=E_{RN}(r_s)=E_s.
\label{bcs}
\end{align}

Using eqs.~(\ref{EME}) and (\ref{f_rn}), we work out
\begin{align}
& f_s=1-\frac{2\kappa_n}{(n-2)A_{n-2}}\frac{M}{r_s^{n-3}}+\frac{4\pi \kappa_n}{(n-2)(n-3)A_{n-2}^2}\frac{Q^2}{r_s^{2(n-3)}},\label{f_s}\\
& \frac{2\pi}{\kappa_n r^2}\left[2(n-3)-2(n-3)f_s+(n-4)r f'(r_s)+r^2 f''(r_s)\right]=\frac{16\pi^2}{A_{n-2}^2}\frac{Q^2}{r_s^{2(n-2)}}\label{f'_s}.
\end{align}

Again considering the equation of boundary position $r_s$, i.e. $\rho_m(r_s)=0$, cf. eq.~(\ref{rho_m}), we simplify eq.~(\ref{f'_s}) to the continuity of the first-order derivative of the metric function at the boundary, $f'(r_s)=f_{RN}'(r_s)=f'_s$. As a result, the parameters $M$ and $Q$ can be expressed in terms of $f_s$ and $f'_s$ as follows,
\begin{align}
M&=\frac{(n-2)A_{n-2}r_s^{n-3}}{\kappa_n}\left[1-f_s-\frac{r_s f'_s}{2(n-3)}\right]\label{M}, \\
Q^2&=\frac{(n-2)(n-3)A_{n-2}^2r_s^{2(n-3)}}{4\pi\kappa_n}\left(1-f_s-\frac{r_s f^{\prime}_s}{n-3}\right)\label{Q}.
\end{align}

\section{Regular Tangherlini-RN black holes under the new interpretation of matter source}
The noncommutative geometry inspired Schwarzschild black hole~\cite{0510112,0807.1939} with the Gaussian-distributed  density is  singularity free at the center and asymptotically Schwarzschild like at large distances ($r \gg \sqrt{\theta}$, see eq.~(\ref{nssf_r})).
However, from the point of view of the casual structure and thermodynamics, it has two horizons and a maximum temperature, behaving more like a Reissner-Nordstr\"{o}m black hole~\cite{1212.5044,0802.1757}. In this section, we construct the regular Tangherlini-RN black hole under the new interpretation of matter source. By starting with the noncommutative geometry inspired Schwarzschild solution in $n$ dimensions, we find that the energy-momentum tensor with the Gaussian distribution can also be regarded as the linear combination of the perfect fluid's and electromagnetic field's energy-momentum tensors when the charge-to-mass ratio takes a special range in order to guarantee the formation of Tangherlini-RN black holes.
In addition, we investigate the stability of the boundary under mass perturbation and indicate that the new interpretation imposes a rigid restriction upon the charge-to-mass ratio. In particular, we observe that such a constraint leads to an interesting corollary that noncommutative black holes
are only stable in the 4-dimensional spacetime.

\subsection{Tangherlini-RN solutions with electromagnetic field}
For the noncommutative Schwarzschild black hole in the $n$-dimensional spacetime, the metric function of eq.~(\ref{metric}) takes the following form,
\begin{align}
f(r)=1-\frac{2k_nm}{(n-2)A_{n-2}\Gamma\left(\frac{n-1}{2}\right)}\frac{\gamma\left(\frac{n-1}{2},\frac{r^2}{4\theta}\right)}{r^{n-3}},
\label{nssf_r}
\end{align}
where $\gamma(s,x)=\int_0^x t^{s-1}e^{-t}dt$ is the lower incomplete gamma function; $\sqrt{\theta}$ is the scale parameter which represents a minimal length, and $m$ the mass parameter.\footnote{For the noncommutative Schwarzschild black hole, the mass distribution is Gaussian, see eq.~(\ref{schrho}). The total amount of this mass parameter is given by $m=A_{n-2}\int_0^\infty \rho r^{n-2} dr$, and the condition that the black hole can be formed is that the average radius of the mass distribution should not be less than the horizon of the hole.}
From the interpretation of the anisotropic fluid, we deduce the density and pressure from eqs.~(\ref{rho})-(\ref{pbot}),
\begin{align}
\rho&=-p_r=\frac{m}{2^{n-2}\theta^{\frac{n-1}{2}}A_{n-2}\Gamma(\frac{n-1}{2})}\exp\left(-\frac{r^2}{4\theta}\right),\label{schrho}\\
p_\bot&=-\rho-\frac{r\rho'}{n-2}=\left[-1+\frac{r^2}{2(n-2)\theta}\right]\rho,
\end{align}
where the Gaussian distribution presented by the density is the result of the noncommutativity of spacetime~\cite{0510112}.
In accordance with the new interpretation of matter source demonstrated in section 2, i.e. effectively from the interpretation of perfect fluid plus electromagnetic field,
we compute the mass density, the isotropic pressure and the energy density of electrostatic field from eqs.~(\ref{rho_m}) and (\ref{EME}),
\begin{align}
\rho_m&=-p=\frac{4(n-2)\theta-r^2}{4(n-2)\theta}\rho,\\
W&=\frac{r^2}{4(n-2)\theta}\rho,
\end{align}
which reveals that the DEC holds within the charged perfect fluid  sphere. Moreover, the boundary of the charged sphere that satisfies $\rho_m=0$  is located at
\begin{align}
r_s=\sqrt{4\theta(n-2)},
\label{nssr_s}
\end{align}
which is of the same order as the scale parameter and grows large when the dimension increases.

Outside the charged perfect fluid sphere, the metric of the electromagnetic vacuum takes the Reissner-Nordstr\"{o}m form, see eq.~(\ref{f_rn}), where the total mass $M$ and charge $Q$ can be calculated from eqs.~(\ref{M}) and (\ref{Q}),
\begin{align}
M&=\frac{m\left[2(n-2)^{\frac{n}{2}}e^{2-n}+(n-3)(n-2)^{\frac{1}{2}}\gamma\left(\frac{n-1}{2},n-2\right)\right]}{(n-3)(n-2)^{\frac{1}{2}}\Gamma\left(\frac{n-1}{2}\right)},\label{M_m}\\
Q^2&=\frac{2^{n-3}(n-2)^{n-2}m A_{n-2}}{\pi e^{n-2} \Gamma\left(\frac{n-1}{2}\right)}\theta^{\frac{n-3}{2}}\label{q^2}.
\end{align}

Eqs.~(\ref{M_m}) and (\ref{q^2}) show that the exterior parameters $M$ and $Q$ are expressed by the interior parameters $m$ and $\theta$ through the boundary.
As a result, the boundary $r_s$ that links the internal  spacetime with the external, can reciprocally be expressed by $M$ and $Q$, or $M$ and the charge-to-mass ratio, $\alpha \equiv Q/M$, as follows,
\begin{align}
r_s^{n-3}=\alpha^2 M\frac{\pi\left[2(n-2)^{\frac{n}{2}}+(n-3)(n-2)^{\frac{1}{2}}e^{n-2}\gamma\left(\frac{n-1}{2},n-2\right)\right]}{(n-3)(n-2)^{\frac{n}{2}}A_{n-2}}.
\label{r_s^n-3}
\end{align}

\begin{figure}[!ht]
\centering\includegraphics[height=7cm]{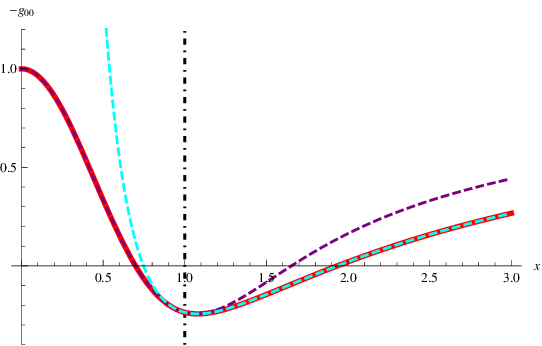}
\caption{Plot of the relation of $-g_{00}$ with respect to the reduced radius $x$ for $n=4$ and ${\alpha^2}/{k_n}=0.032$. The purple and cyan dashed curves correspond to $f(x)$ and $f_{RN}(x)$, respectively, and they intersect at $x=1$ where the black dot-dashed line lies. The red curve describes $f(x)$ for $x\leq1$ and $f_{RN}(x)$ for $x>1$.}
\label{-g_00}
\end{figure}

Further using eq.~(\ref{r_s^n-3}) together with eqs.~(\ref{M_m}) and (\ref{q^2}), we reduce the interior metric eq.~(\ref{nssf_r}) and the exterior metric eq.~(\ref{f_rn}) to the following forms,
\begin{align}
f(x)\!=\!1\!-\!\frac{2k_n}{\pi\alpha^2} \frac{(n-3)^2(n-2)^{\frac{n-1}{2}}e^{n-2}}{\left[2(n-2)^{\frac{n}{2}}\!+\! (n-3)(n-2)^{\frac{1}{2}}e^{n-2}\gamma\left(\frac{n-1}{2},n-2\right)\right]^2}\frac{\gamma\left(\frac{n-1}{2},(n-2)x^2\right)}{x^{n-3}}, \label{f(x)}
\end{align}
and
\begin{align}
f_{RN}(x)
=1-\frac{2k_n}{\pi\alpha^2}\Bigg\{\frac{(n-3)(n-2)^{\frac{n}{2}-1}}
{2(n-2)^{\frac{n}{2}}+(n-3)(n-2)^{\frac{1}{2}}e^{n-2}\gamma\left(\frac{n-1}{2},n-2\right)}\frac{1}{x^{n-3}} \nonumber \\
-\frac{2(n-3)(n-2)^{n-1}}{\left[2(n-2)^{\frac{n}{2}}+(n-3)
(n-2)^{\frac{1}{2}}e^{n-2}\gamma\left(\frac{n-1}{2},n-2\right)\right]^2}\frac{1}{x^{2(n-3)}}\Bigg\} \label{f_rn(x)},
\end{align}
where $x$, the reduced radius, is defined by $x\equiv{r}/{r_s}$. It is obvious that on the boundary, i.e. at $x=1$, $f(x)$ equals naturally to $f_{RN}(x)$. The two metrics are illustrated in Fig.~\ref{-g_00} for a specific dimension and charge-to-mass ratio.
Eq.~(\ref{f(x)}) indicates that the exterior parameters ($M$ and $\alpha$) can reflect the interior structure, so next we analyze how the value of $\alpha$ affects the geometrical characters in both the interior and the exterior of the  charged perfect fluid sphere.

Before our discussions, we have to mention the prerequisite that a system with certain mass and charge forms a black hole. A  hypothesis proposed by Thorne~\cite{thorne} states that a black hole forms only when the spherical radius of the system is not larger than its Schwarzschild radius, known as the hoop conjecture. Similarly, the hoop conjecture of the charged case was put forward~\cite{1511.03665}, meaning that a physical system of mass $M$ and electric charge $Q$ forms a black hole if its circumference radius $r_c$ is not larger than the corresponding RN black hole radius,\footnote{In the 4-dimensional spacetime, $k_n=8\pi$ has been considered.} $r_{RN}=M+\sqrt{M^2-Q^2}$. Based on this conjecture, one has to distinguish~\cite{0612035} the two basic categories: A black hole with horizons and a massive charged ``droplet", both of which have a finite-sized charged perfect fluid sphere of the radius $r_s$. Now we study the geometrical characters in both the interior and the exterior of the charged perfect fluid sphere in detail by considering the following three cases.

\subsubsection{Case I}
Setting $f_{RN}(r)$ of eq.(\ref{f_rn}) be zero, we locate the horizons of the RN black hole at
\begin{align}
r_{\pm}^{n-3}=\frac{k_n M}{(n-2)A_{n-2}}\left(1\pm\sqrt{1-\frac{4\pi(n-2)}{k_n(n-3)}\alpha^2}\right),\label{rnrpm}
\end{align}
under the condition
\begin{align}
\beta \equiv \frac{\alpha^2}{k_n} \leq \beta_1\equiv \frac{1}{4\pi}\left(\frac{n-3}{n-2}\right),
\end{align}
where $\beta$ denotes a reduced charge-to-mass ratio squared. On the contrary, if $\beta>\beta_1$, then no real roots of $f_{RN}(r)=0$ exist, meaning that no horizons exist to the RN metric and further indicating that no black holes form according to the hoop conjecture. So $\beta_1$ is the upper bound for the existence of a black hole, whose possible values are listed in the first line of Table~\ref{table_bound} for various dimensions.

\subsubsection{Case II}
If horizons exist in the RN metric, i.e. $\beta\leq\beta_1$ is satisfied, the position of boundary relative to the RN horizons needs to be fixed, which generally leads to the three subcases of the hoop conjecture: $r_s<r_-$, $r_-<r_s<r_+$, and $r_s=r_-$ or $r_s=r_+$.\footnote{In the particular situation of the extremal black hole at $\beta=\beta_1$, the classification is still effective as long as the additional condition $r_-=r_+$ is considered in the subcases.}

$\bullet$ \emph{$r_s< r_-$ shows that the boundary is hidden behind the inner horizon}

Using eq.~(\ref{rnrpm}), we rewrite this condition to be
\begin{equation}
r_s^{n-3}<r_-^{n-3}=\frac{k_n M}{(n-2)A_{n-2}}\left(1-\sqrt{1-\frac{4\pi(n-2)}{n-3}\beta}\right),
\label{r_s^<r_-^}
\end{equation}
which gives equivalently the following constraint to $\beta$,
\begin{equation}
\beta_2<\beta<\beta_3,
\end{equation}
where the newly introduced parameters are defined by
\begin{align}
\beta_2&\equiv\frac{2(n-3)^2(n-2)^{\frac{n-1}{2}}e^{n-2}\gamma\left(\frac{n-1}{2},n-2\right)}
{\pi \left[2(n-2)^{\frac{n}{2}}+(n-3)(n-2)^{\frac{1}{2}}e^{n-2}\gamma\left(\frac{n-1}{2},n-2\right)\right]^2},\label{beta_2}\\
\beta_3&\equiv\frac{(n-3)(n-2)^{\frac{n}{2}-1}}{\pi \left[2(n-2)^{\frac{n}{2}}+(n-3)(n-2)^{\frac{1}{2}}e^{n-2}\gamma\left(\frac{n-1}{2},n-2\right)\right]}.\label{beta_3}
\end{align}

\begin{figure}[!ht]
\centering\includegraphics[height=13cm]{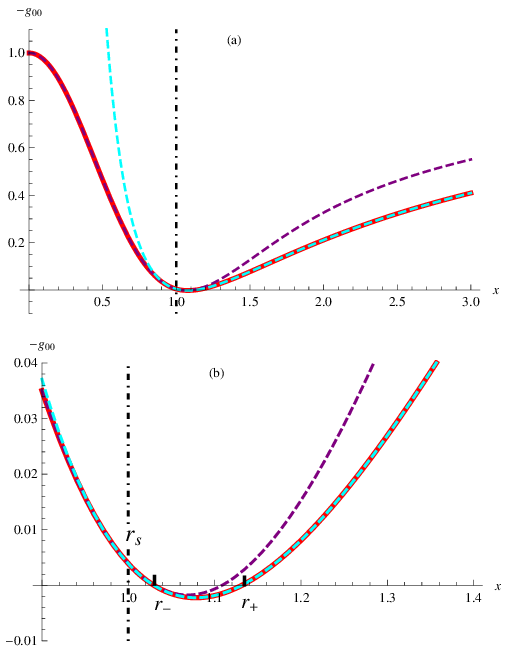}
\caption{An example of $\beta\leq\beta_1$ and $\beta_2<\beta<\beta_3$: $n=4$, $\beta=0.0397$, where Fig.~(b) is the magnified image of Fig.~(a) within the range $x\in[0.9, 1.3]$. This case is characterized by $r_s<r_-$ and $f(x)>0$ for $x\in [0,1]$.}
\label{r_s<r_-}
\end{figure}

We list $\beta_2$ and $\beta_3$ in the second and third rows of Table~\ref{table_bound}, respectively, for various dimensions. From the data, we can see that the two conditions, $\beta\leq\beta_1$ and $\beta_2<\beta<\beta_3$, can be held simultaneously only in the 4-dimensional spacetime, where the first condition guarantees the formation of the black hole and the second keeps the interior of the black hole, i.e. the charged perfect fluid sphere, hidden behind the inner horizon. For illustration, see Fig.~$\ref{r_s<r_-}$.

\begin{figure}[!ht]
\centering\includegraphics[height=7cm]{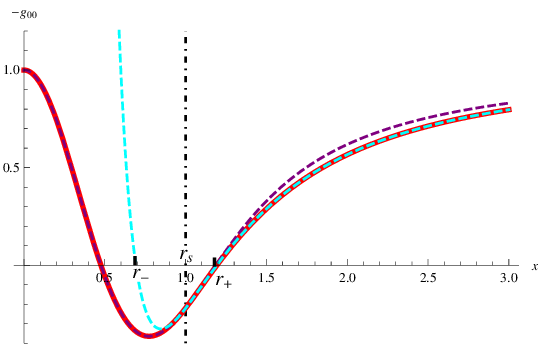}
\caption{An example of $\beta<\beta_1$ and $\beta<\beta_2$: $n=5$, $\beta=0.04$. This case is characterized by $r_-<r_s<r_+$ and one $f(r)$-horizon.}
\label{r_-<r_s<r_+}
\end{figure}

\begin{figure}[!ht]
\centering\includegraphics[height=7cm]{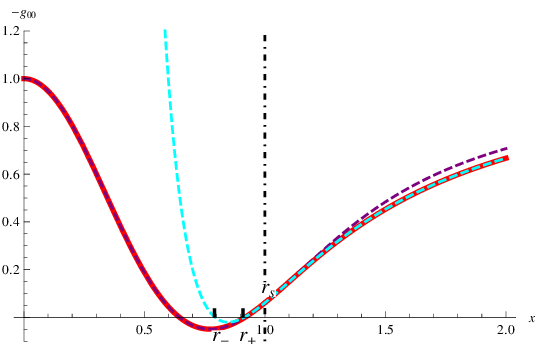}
\caption{An example of $\beta\leq\beta_1$, $\beta>\beta_2$ and $\beta>\beta_3$: $n=5$, $\beta=0.052$. This case is characterized by $r_s>r_+$.}
\label{r_s>r_+}
\end{figure}

$\bullet$ \emph{$r_-<r_s< r_+$ shows that the boundary is not hidden behind the inner horizon but the event horizon}\footnote{To avoid ambiguities, the inner horizon and event horizon specially refer to the two roots of $f_{RN}(r)$ in the present paper, $r_-$ and $r_+$. As to the roots of $f(r)$, we call them the ``$f(r)$-horizons" that are not larger than $r_+$.}

The inequality $r_-<r_s<r_+$ leads to $\beta<\beta_2$. Moreover, it should be noted that $r_-<r_s$ does not mean that the inner horizon of RN metric hides behind the boundary. In fact, the whole interior of the charged perfect fluid sphere is described by the metric function $f(r)$, not by $f_{RN}(r)$ that refers to the exterior of the charged sphere. Thus, $\beta<\beta_2$, together with $\beta<\beta_1$, describes the situation that the boundary lies behind the event horizon, but not the inner horizon. This situation is possible since $\beta_2<\beta_1$ is tenable for all dimensions, see Fig.~$\ref{r_-<r_s<r_+}$ for instance.

$\bullet$ \emph{$r_s=r_-$ or $r_s=r_+$ shows that the boundary coincides with one of the $f_{RN}(r)$ horizons}

Our calculations show that $r_s=r_-$ requires the constraints $\beta=\beta_2$ and $\beta\leq\beta_3$, and that $r_s=r_+$ requires $\beta=\beta_2$ and $\beta\geq\beta_3$. From the data in Table~\ref{table_bound}, we can deduce that $\beta=\beta_2$ represents the case that the boundary $r_s$ coincides either with the inner horizon $r_-$ for $n=4$, or with the event horizon $r_+$ for $n\geq5$.

At the end of Case II, we clarify the meaning of $r_s>r_+$. This inequality corresponds to the constraints, $\beta>\beta_2$ and $\beta>\beta_3$. As we have explained above for $r_-<r_s$,  the situation $r_s>r_+$ does not imply that the event horizon of RN metric hides behind the boundary.
It depicts a finite-sized charged perfect fluid sphere with no RN horizons outside the boundary, as was mentioned above by the second category -- a massive charged ``droplet". For illustration, see Fig.~$\ref{r_s>r_+}$.

\subsubsection{Case III}
Here we revisit $r_s\leq r_+$ on how $f(r)$ (or $f(x)$) determines the structure of the interior of the charged perfect fluid sphere and how it affects the characters of the black hole. To be specific, if the restriction $f(x)>0$ is obeyed within $x\in[0,1]$, no horizons can be formed by $f(r)$. $f(r)$ horizons appear if and only if $f(r)=0$. 
We compute the roots of $f'(x_*)=0$ from $n=4$ to $11$ dimensions,
\begin{equation}
x_*=1.06859, \quad 0.77315, \quad 0.62760, \quad 0.53874, \quad 0.47797, \quad 0.43331, \quad 0.39885, \quad 0.37128.
\end{equation}

The data show that there is only one minimum for $f(x)$. So $f(x)>0$ for all $x\in[0,1]$ corresponding to no $f(r)$ horizons, is equivalent to
\begin{align}
f(1)>0, \quad\text{for}\,\, n=4; \qquad f(x_*)>0, \quad \text{for}\,\, n=5\,\,\text{to}\,\, 11,
\end{align}
which gives $\beta$ the constraint,
\begin{equation}
\beta>\beta_4,
\end{equation}
with the definition of $\beta_4$ as follows,
\begin{align}
\beta_4\equiv\frac{4\sqrt{2}e^2\gamma\left(\frac{3}{2},2\right)}{\pi\left[8+\sqrt{2}e^2\gamma\left(\frac{3}{2},2\right)\right]^2}&, \qquad \qquad \,\text{for}\,\, n=4; \notag\\
\beta_4\equiv\frac{2}{\pi} \frac{(n-3)^2(n-2)^{\frac{n-1}{2}}e^{n-2}}{\left[2(n-2)^{\frac{n}{2}}\!+\! (n-3)(n-2)^{\frac{1}{2}}e^{n-2}\gamma\left(\frac{n-1}{2},n-2\right)\right]^2}&\frac{\gamma\left(\frac{n-1}{2},(n-2)x_*^2\right)}{x_*^{n-3}},
\,\text{for}\,\, n\!=\!5\,\,\text{to}\,\, 11.
\end{align}

 The values of $\beta_4$ are listed in the fourth row of Table~\ref{table_bound} for various dimensions. From the data, we can see that the two conditions, $\beta>\beta_4$ and $\beta\leq\beta_1$, which guarantee a black hole with no $f(r)$ horizons, are only compatible in the 4-dimensional spacetime. For illustration,  see Fig.~\ref{r_s<r_-} that reflects the feature: $f(r)>0$, $x\in[0,1]$.

On the contrary, if $\beta<\beta_4$, $f(r)$ horizons would possibly form due to the existence of the roots of $f(x)=0$ within $x\in[0,1]$. As mentioned before, only one minimum exists for $f(x)$, so there are at most two roots for $f(x)=0$. Moreover, the number of roots within $x\in[0,1]$ depends on the sign of the metric function at the boundary, i.e. eq.~(\ref{f(x)}) at $x=1$. That is, there is one root of $f(x)=0$ for $f(1)<0$ and there are two roots for $f(1)>0$.

Referring to eq.~(\ref{f(x)}), we obtain that the inequality $f(1)<0$, i.e.
\begin{align}
f(1)\!=\!1\!-\!\frac{2}{\pi\beta} \frac{(n-3)^2(n-2)^{\frac{n-1}{2}}e^{n-2}}{\left[2(n-2)^{\frac{n}{2}}\!+\! (n-3)(n-2)^{\frac{1}{2}}e^{n-2}\gamma\left(\frac{n-1}{2},n-2\right)\right]^2}\gamma\left(\frac{n-1}{2},(n-2)\right)<0 \label{f(1)}
\end{align}
gives exactly the constraint $\beta<\beta_2$, cf. eq.~(\ref{beta_2}). As $\beta<\beta_2$ corresponds to the second subcase of Case II, $r_-<r_s<r_+$, the only root of $f(x)=0$ within $x\in[0,1]$ is the reduced radius of the $f(r)$ horizon. See Fig.~\ref{r_-<r_s<r_+} for instance.

In contrast, the inequality $f(1)>0$ leads to $\beta>\beta_2$. Although there are two roots of $f(x)=0$ within $x\in[0,1]$, the condition $\beta_2<\beta<\beta_4$ implies the satisfaction of $\beta>\beta_2$ and $\beta>\beta_3$ for $n\geq5$ from the data in Table~\ref{table_bound}, which conforms to the situation $r_s>r_+$ that fails to form a black hole. For illustration, see Fig.~\ref{r_s>r_+}.

Particularly, $\beta=\beta_2$ is obtained from solving $f(1)=0$, which is consistent with the third subcase of Case II, $r_s=r_-$ or $r_s=r_+$. So we can infer that for $\beta=\beta_2$ and $n=4$, $r_s=r_-$ is the radius of the only horizon formed by $f(r)$ as a result of $\beta_2=\beta_4$, and that for $\beta=\beta_2$ and $n\geq5$, in addition to $r_s=r_+$, there exists another $f(r)$ horizon inside the sphere due to $\beta_2<\beta_4$.

\begin{table}[h]
\centering
\begin{tabular}{|m{2.15cm}*{8}{|c}|}
\hline
\multirow{2}*{Restrictions} & \multicolumn{8}{c|}{Dimensions of spacetime}  \\ \cline{2-9}
                   &  4 &  5  & 6 & 7 & 8 & 9 & 10 & 11 \\
\hline

            $\beta_1$  &  0.03979  & 0.05305   & 0.05968   & 0.06366  &  0.06631  &   0.06820 &  0.06963 & 0.07074 \\
\hline

            $\beta_2$  &  0.03955 &  0.04882 & 0.04574 & 0.03980 & 0.03369 & 0.02818 & 0.02345 & 0.01947 \\

\hline
            $\beta_3$  &  0.04290 &  0.03807 & 0.03084 & 0.02469 & 0.01980 & 0.01596 & 0.01292 & 0.01052 \\

\hline
            $\beta_4$  &  0.03955  & 0.05458   & 0.07090   & 0.09662  &  0.1384  &   0.2065 &  0.3187 & 0.5053 \\
\hline
\end{tabular}
\caption{The restrictions on the reduced charge-to-mass ratio squared, $\beta\equiv {\alpha}^2/k_n$, in various dimensions. }
\label{table_bound}

\end{table}

With all factors taken into account, we categorize the black hole solutions with a finite-sized charged perfect fluid sphere as the matter source by the following situations:

$\mathrm{(\rmnum{1})}$ The charged perfect fluid with positive metric function $f(r)$ is hidden inside the inner horizon of the RN metric when the inequality, $\beta_2<\beta\leq\beta_1$, is satisfied in the 4-dimensional spacetime;  

$\mathrm{(\rmnum{2})}$ The charged perfect fluid with one root of $f(r)=0$ within $r\in[0,r_s]$ and negative boundary metric function $f(r_s)$ is hidden inside the event horizon but not the inner horizon of the RN metric when the inequality, $\beta<\beta_2$, is satisfied in all dimensions;  

$\mathrm{(\rmnum{3})}$ The charged perfect fluid takes a vanishing boundary metric function $f(r_s)=0$ when $\beta=\beta_2$ is satisfied, specifically, the boundary coincides with the inner horizon of the RN metric at $n=4$ as the only root of $f(r_s)=0$, and with the event horizon in the other dimensions as the second root.

\subsection{Stability of the boundary}
The possible regular black hole solutions have been obtained in all dimensions if the hoop conjecture is considered. However, from the point of view of stability, the stable configuration for the noncommutative model in dimensions higher than 4 ($n\geq5$) would be a massive charged ``droplet" rather than a regular black hole since the outer layers of the black hole with negative metric function $f(r)$ could not stay in a static equilibrium~\cite{1706.03454}. As a result, only the 4-dimensional spacetime seemingly admits stable regular black holes with a static interior spacetime structure for $\beta_2\leq\beta\leq\beta_1$ due to the positivity of the internal metric function $f(r)$. Nonetheless, the stability of the boundary still remains to be analyzed since the discontinuities and non-smoothness of the boundary possibly happen.

As we have known in subsection 2.3, the boundary conditions (eq.~(\ref{bcs})) are equivalent to the continuity of the metric function and its first-order derivative, and thus they guarantee the continuity of the extrinsic curvature and satisfy Israel's condition~\cite{Israel} about two spacetimes' soldering,
\begin{align}
K_{ab}\mid^+=K_{ab}\mid^-,
\end{align}
where $K_{ab}$ is the extrinsic curvature defined with the induced metric $h_{ab}$ and the normal vector $n^a$ of the boundary hypersurface $\Sigma$ as follows,
\begin{align}
K_{ab}:=h_a^{\,\, c}h_b^{\,\,d}\nabla_c n_d,
\label{K_ab}
\end{align}
$K_{ab}\mid^+$ and $K_{ab}\mid^-$ denote the limit of $K_{ab}$ on the boundary $\Sigma$ from the exterior spacetime ($V^+$) and the interior spacetime ($V^-$), respectively.
Nevertheless, in the specific case, $\beta_2\leq\beta\leq\beta_1$ and $n=4$, the boundary coincides with ($\beta=\beta_2$) or hides behind ($\beta_2<\beta\leq\beta_1$) the inner horizon of the RN metric marked with the ``mass inflation" phenomenon~\cite{mass inflation}, that is, the inner horizon becomes unstable under perturbations. As a result, we exclude the lightlike boundary for $\beta=\beta_2$ from consideration, and adopt an easier way to check whether the timelike boundary for $\beta_2<\beta\leq\beta_1$ keeps stable or not under mass perturbations~\cite{0804.0295,1209.3567}.

For the convenience in the following discussions, we start with rewriting the interior and the exterior metrics of the black hole, see eq.~(\ref{nssf_r}) and eq.~(\ref{f_rn}), in the following unified form,
\begin{align}
\mathrm{d}s^2=-h(r)\mathrm{d}t^2+\frac{1}{h(r)}\mathrm{d}r^2+r^2\,\mathrm{d}\Omega_{n-2}^2,
\end{align}
with
\begin{align}
h(r)=\Bigg\{\begin{array}{lcl}
h_i(r)=f(r)=1-\frac{2k_n}{(n-2)A_{n-2}}\frac{m}{r^{n-3}\Gamma(\frac{n-1}{2})}\gamma(\frac{n-1}{2},\frac{r^2}{4\theta}), \quad \text{for}\,\, r<R,\\
h_e(r)=f_{RN}(r)=1-\frac{2\kappa_n}{(n-2)A_{n-2}}\frac{M}{r^{n-3}}+\frac{4\pi \kappa_n}{(n-2)(n-3)A_{n-2}^2}\frac{Q^2}{r^{2(n-3)}}, \quad \text{for}\,\, r>R,\label{hihe}
\end{array}
\end{align}
where we re-label the boundary that separates two spacetimes with $R$ to distinguish it from $r_s$ in the non-perturbative case.
Then the induced metric on the boundary $\Sigma$ can be written as
\begin{align}
(\mathrm{d}s^2)_\Sigma=-\mathrm{d}\tau^2+R^2(\tau)\,\mathrm{d}\Omega_{n-2}^2,
\end{align}
where $\tau$ is the proper time and $R(\tau)$ represents the evolution of the boundary.
Under the precondition that we discuss the influence of a mass perturbation on the timelike boundary, the energy-momentum tensor of matter on $\Sigma$ (or of the massive shell) is assumed to take the following form,
\begin{align}
S_{ab}=\sigma(\tau)v_a v_b,
\end{align}
where $\sigma(\tau)$ is the surface energy density and $v^a$ the normalized velocity of the comoving observer on $\Sigma$.  We use the normalized velocity in its following form,
\begin{align}
v^a=\left(\frac{\partial}{\partial \tau}\right)^a=\dot{t}\left(\frac{\partial}{\partial t}\right)^a+\dot{R}\left(\frac{\partial}{\partial R}\right)^a,
\end{align}
where the dot denotes the derivative with respect to $\tau$ and $\dot{t}$ can be expressed in terms of $h(R)$ and $\dot{R}$ from $g_{ab}v^av^b=-1$ as follows,
\begin{align}
\dot{t}=\frac{\sqrt{h(R)+\dot{R}^2}}{h(R)}.
\end{align}

Further, we derive the normal vector $n^a$ from $n^av_a=0$ and $n^an_a=1$ to be
\begin{align}
n^a=n^t\left(\frac{\partial}{\partial t}\right)^a+n^r\left(\frac{\partial}{\partial R}\right)^a=\frac{\dot{R}}{h(R)}\left(\frac{\partial}{\partial t}\right)^a+\sqrt{h(R)+\dot{R}^2}\left(\frac{\partial}{\partial R}\right)^a,
\end{align}
where the direction of $n^a$ is set to point from $V^-$ to $V^+$.

Applying the energy conversation law to $S_{ab}$~\cite{0804.0295}, that is, $^{n-1}\nabla^a S_{ab}=0$, with $^{n-1}\nabla_a$ denoting the covariant derivative
on the $(n-1)$-dimensional hypersurface, we obtain
\begin{align}
\frac{\mathrm{d}(R^{n-2}\sigma)}{\mathrm{d}\tau}=0,
\end{align}
which indicates that we can define a proper mass of the shell that is independent of time $\tau$,
\begin{align}
M_s=A_{n-2}\sigma R^{n-2}.
\label{M_s}
\end{align}

Moreover, as a result of the existence of $S_{ab}$, the extrinsic curvature undergoes a jump on the boundary and satisfies~\cite{Israel}
\begin{align}
[K_{ab}]-h_{ab}[K]=-\kappa_n S_{ab},
\label{K_ab1}
\end{align}
where $K=h_{ab}K^{ab}$ is the trace and the brackets denote the difference between the limits of quantity on $\Sigma$ from $V^+$ to $V^-$, e.g. $[K_{ab}]=K_{ab}\mid^+-K_{ab}\mid^-$ and $[K]=K\mid^+-K\mid^-$. For convenience, we write eq.~(\ref{K_ab1}) in another form,
\begin{align}
[K_{ab}]=-\kappa_n\left(S_{ab}-\frac{1}{n-2}h_{ab}S\right),
\label{K_ab2}
\end{align}
where $S=h^{ab}S_{ab}=-\sigma$. From eq.~(\ref{K_ab}), we derive one of the nonvanishing components,
\begin{align}
K_{\theta\theta}=R n^r=R\sqrt{h(R)+\dot{R}^2}.
\label{K_theta}
\end{align}

Further using eq.~(\ref{K_theta}) together with eqs.~(\ref{M_s}) and (\ref{K_ab2}), we obtain
\begin{align}
[K_{\theta\theta}]=R\left(\sqrt{h_e(R)+\dot{R}^2}-\sqrt{h_i(R)+\dot{R}^2}\right)=-\frac{k_n}{n-2}\sigma R^2=-\frac{k_n M_s}{(n-2)A_{n-2}R^{n-4}},
\label{k_theta}
\end{align}
which links the jump of the metric function on the boundary $\Sigma$ with the shell mass $M_s$.
Using eq.~(\ref{hihe}), we can see that eq.~(\ref{k_theta}) is in fact the equation of motion of the massive shell~\cite{1209.3567},
\begin{align}
\dot{R}^2+V(R)=-1,
\label{motion}
\end{align}
where
\begin{align}
V(R)\!=\!-\left(\frac{2\pi Q^2}{(n-3)A_{n-2}R^{n-3}M_s}\!-\!\frac{M-m(R)}{M_s}\!-\!\frac{k_n M_s}{2(n-2)A_{n-2}R^{n-3}}\right)^2\!-\!\frac{2k_n m(R)}{(n-2)A_{n-2}R^{n-3}}\label{vform}
\end{align}
can be regarded as the potential function, and $m(R)\equiv\frac{m}{\Gamma(\frac{n-1}{2})}\gamma(\frac{n-1}{2},\frac{R^2}{4\theta})$. Based on that the timelike boundary only exists in the 4-dimensional spacetime and $M_s$ is a perturbation ($M_s\rightarrow0$), we now focus on eq.~(\ref{vform}) for $n=4$,
\begin{align}
V(R)=-\frac{[Q^2-2MR+2m(R)R]^2}{4M_s^2 R^2}-\frac{2m(R)}{R}.
\label{v(R)}
\end{align}

Then, a stable stationary solution of motion would reasonably satisfies
\begin{align}
\dot{R}=&0 \quad \Leftrightarrow \quad V(R)=-1
\label{v(r)}\\
\frac{dV(R)}{dR}=&\frac{[Q^2-2m'(R)R^2][Q^2-2MR+2m(R)]}{2M_s^2 R^3}-\frac{2m'(R)R-2m(R)}{R^2}=0,
\label{dv/dr}\\
\frac{d^2V(R)}{dR^2}=&-\frac{[Q^2-2m'(R)R^2]^2+2[Q^2-2MR+2m(R)R][m''(R)R^3+Q^2]}{2M_s^2 R^4} \notag \\
&-\frac{2m''(R)R^2-4m'(R)R+4m(R)}{R^3}>0.
\end{align}

Similarly, applying the setting $M_s\rightarrow0$ to eqs.~(\ref{v(R)}) and (\ref{dv/dr}), we derive
\begin{align}
Q^2-2MR+2m(R)R=\mathcal{O}(M_s),\\
Q^2-2m'(R)R^2=\mathcal{O}(M_s),
\end{align}
which keep $V(R)$ and $V'(R)$ finite. Henceforth, the parameters $Q$ and $M$ in the outer RN metric are related to the interior parameters, $m$ and $\theta$, and the shell mass $M_s$ as
\begin{align}
Q=&\sqrt{\frac{mR^4\exp\left(-\frac{R^2}{4\theta}\right)}{\pi^{\frac{1}{2}}\theta^{\frac{3}{2}}}}-A\exp\left(\frac{R^2}{8\theta}\right)M_s+\mathcal{O}(M_s^2),\label{Q1}\\
M=&\frac{2m}{\pi^{\frac{1}{2}}}\gamma\left(\frac{3}{2},\frac{R^2}{4\theta}\right)+\frac{mR^3\exp\left(-\frac{R^2}{4\theta}\right)}{2\pi^{\frac{1}{2}}\theta^{\frac{3}{2}}}
  +B\sqrt{\frac{m}{\pi^{\frac{1}{2}}\theta^{\frac{1}{2}}}}M_s+\mathcal{O}(M_s^2), \label{M1}
\end{align}
where $A$ and $B$ are two coefficients to be determined. If we take $M_s=0$, $Q$ and $M$ spontaneously go back to eqs.~(\ref{q^2}) and (\ref{M_m}) for $n=4$, respectively.
Combining eqs.~(\ref{Q1}) and (\ref{M1}) with eqs.~(\ref{v(r)}) and (\ref{dv/dr}), the parameters $A$ and $B$ can be determined by
\begin{align}
\frac{m}{\pi^{\frac{1}{2}}\theta^{\frac{1}{2}}}\left(\frac{R_s}{\theta^{\frac{1}{2}}}A+B\right)^2&
=1-\frac{4m}{\pi^{\frac{1}{2}}R_s}\gamma\left(\frac{3}{2},\frac{R_s^2}{4\theta}\right),\label{root_1}\\
\frac{2m}{\pi^{\frac{1}{2}}\theta}\left(\frac{R_s}{\theta^{\frac{1}{2}}}A^2+AB\right)&=\frac{1}{R_s^2}
\left[\frac{mR_s^3\exp\left(-\frac{R_s^2}{4\theta}\right)}{\pi^{\frac{1}{2}}\theta^{\frac{3}{2}}}
-\frac{4m}{\pi^{\frac{1}{2}}}\gamma\left(\frac{3}{2},\frac{R_s^2}{4\theta}\right)\right],\label{root_2}
\end{align}
where $R_s$ is the radius of the stationary solution in the limit of $M_s\rightarrow 0$, that is, $R_s=r_s$.
Now we know from the above relevant equations that $\frac{d^2V(R)}{dR^2}$ is approximately equal to
\begin{align}
\frac{d^2V(R)}{dR^2}\approx&\frac{mR_s\left(R_s^2-2\theta \right)\exp\left(-\frac{R_s^2}{4\theta}\right)}{2\pi^{\frac{1}{2}}\theta^{\frac{5}{2}}M_s}\sqrt{\frac{m}{\pi^{\frac{1}{2}}\theta^{\frac{1}{2}}}}\left(\frac{R_s}{\theta^{\frac{1}{2}}}A+B\right).
\label{d^2v}
\end{align}

Substituting the solution $R=R_s$ into eq.~(\ref{k_theta}), we have
\begin{align}
&\sqrt{h_e(R_s)+\dot{R_s}^2}-\sqrt{h_i(R_s)+\dot{R_s}^2} \notag \\
=&\sqrt{1-\frac{4m}{\pi^{\frac{1}{2}}R_s}\gamma\left(\frac{3}{2}, \frac{R_s^2}{4\theta}\right)-\frac{1}{R_s}\sqrt{\frac{m}{\pi^{\frac{1}{2}}\theta^{\frac{1}{2}}}}\left(\frac{R_s}{\theta^{\frac{1}{2}}}A+B\right)M_s}-\sqrt{1-\frac{4m}
{\pi^{\frac{1}{2}}R_s}\gamma\left(\frac{3}{2}, \frac{R_s^2}{4\theta}\right)} \notag \\
=&-\frac{k_n M_s}{(n-2)A_{n-2}R_s^{n-3}}.
\end{align}

We infer that $\left(\frac{R_s}{\theta^{\frac{1}{2}}}A+B\right)$ takes the positive root of eq.~(\ref{root_1}) and further obtain the second-order derivative of $V(R)$ (eq.~(\ref{d^2v})) as
\begin{align}
\frac{d^2V(R)}{dR^2}\approx&\frac{mR_s\left(R_s^2-2\theta \right)\exp\left(-\frac{R_s^2}{4\theta}\right)\sqrt{1-\frac{4m}{\pi^{\frac{1}{2}}R_s}\gamma\left(\frac{3}{2},\frac{R_s^2}{4\theta}\right)}}{2\pi^{\frac{1}{2}}\theta^{\frac{5}{2}}M_s},
\end{align}
which is positive for a physical perturbation $M_s>0$ since $R_s=r_s=\sqrt{8\theta}$ and $f(R_s)=1-\frac{4m}{\pi^{\frac{1}{2}}R_s}\gamma\left(\frac{3}{2},\frac{R_s^2}{4\theta}\right)>0$. Therefore, we can conclude that the timelike boundary is stable under a positive mass perturbation. Table~\ref{tabel_example}, for instance, gives some data which coincide with the conclusion.\footnote{In subsection 3.1, we have obtained the range of $\beta$ for the 4-dimensional black hole with timelike boundary, $\beta\in (0.03955,0.03979)$. It is actually equivalent to the range of the interior parameter, $\frac{m}{\sqrt{\theta}}\in(1.90318, 1.91489)$.}

\begin{table}[h]
\centering
\begin{tabular}{|c|c|c|c|c|c|}
\hline
                    $m/\sqrt{\theta}$ &  $Q_0/\sqrt{\theta}$  & $M_0/\sqrt{\theta}$  &  $(Q-Q_0)/\sqrt{\theta}$ &  $(M-M_0)/\sqrt{\theta}$  & $\frac{d^2V(R)}{dR^2}\theta$ \\
\hline
\hline

            1.905  &  3.051 & 3.053  & $-2.960\times10^{-5} $  & $-1.955\times10^{-5}$ &  $ 3.464\times10^3 $  \\
\hline

           1.908  &  3.054 & 3.057  & $-3.549\times10^{-5} $  & $-2.392\times10^{-5}$ &  $ 2.895\times10^3 $  \\

\hline
           1.911  &  3.056 & 3.062  & $-4.728\times10^{-5} $  & $-3.251\times10^{-5}$ &  $ 2.179\times10^3 $  \\

\hline
            1.914  &  3.058 & 3.067  & $-9.899\times10^{-5} $  & $-6.941\times10^{-5}$ &  $ 1.043\times10^3 $  \\
\hline
\end{tabular}
\caption{The values of physical quantities varied with $m$: $Q_0$ and $M_0$ at $M_s=0$, and the deviations $Q-Q_0$ and $M-M_0$, and the second-order derivative $\frac{d^2V(R)}{dR^2}$ at $M_s=10^{-5}$. }
\label{tabel_example}

\end{table}

\section{Summary}
In this paper, we generalize the formulation that a regular RN black hole can be sourced from charged perfect fluid in the 4-dimensional spacetime to higher-dimensional spacetimes through decomposing the energy-momentum tensor of an anisotropic fluid effectively into the energy-momentum tensor of a perfect fluid plus linear electromagnetic field. Under this new interpretation of matter source, the dominant energy condition is guaranteed as in the spacial case of the 4-dimensional spacetime~\cite{1706.03454}.
In addition, the matter source has a natural boundary $r_s$ where the density and pressure of the perfect fluid vanish and so they no longer occupy the entire spacetime.\footnote{The density and pressure occupy the entire spacetime in the interpretations associated with anisotropic fluid~\cite{ 0801.3519,1511.00853}.} Consequently, the interior of the black hole naturally connects to the exterior electrovacuum through the boundary. As a result, the interior structure is revealed by the parameters of the black hole -- the mass and the charge, by means of the boundary conditions.

We investigate the model with the noncommutative geometry inspired Schwarzschild solution as the interior solution under the new interpretation of matter source. The interior structure is analyzed at three different levels: 

$\mathrm{(\rmnum{1})}$ If the hoop conjecture is satisfied ($r_s\leq r_+$), such a distribution would form a black hole; 

$\mathrm{(\rmnum{2})}$ The boundary hides behind or lies at the inner horizon only if $r_s\leq r_-$, or else the inner horizon does not exist; 

$\mathrm{(\rmnum{3})}$ No horizons are formed by $f(r)$ if the interior metric function $f(r)>0$ stands within the whole region of $r<r_s$, otherwise $f(r)$ horizons would exist. 

The classification is characterized by the value of the charge-to-mass ratio $\alpha$ (or the reduced parameter $\beta$). In this specific model, the possible interior structure of black hole solutions can be summarized as: 

$\mathrm{(\rmnum{1})}$ The matter source with positive metric function $f(r)$ hides inside the inner horizon and no $f(r)$ horizons form in the 4-dimensional spacetime; 

$\mathrm{(\rmnum{2})}$ The matter source with negative $f(r)$ at the boundary hides inside the event horizon, but not inside the inner horizon and one $f(r)$ horizon forms in all the dimensional spacetimes; 

$\mathrm{(\rmnum{3})}$ The matter source with a vanishing $f(r)$ at the boundary has only one $f(r)$ horizon that coincides with the inner horizon in the $4$-dimensional spacetime, and has two $f(r)$ horizons, of which the second one coincides with the event horizon in the other dimensional spacetimes.

When the stability of boundary is taken into consideration, we focus on the black hole solution in the 4-dimensional spacetime, whose boundary that connects two different spacetimes is a timelike hypersurface. By investigating the equation of motion of the shell, we show that this shell is stationary and stable under a mass perturbation. Thus, the noncommutative model permits stable black hole solutions only in the 4-dimensional spacetime under the new interpretation, which gives a novel restriction on dimensions that was not mentioned in a previous work~\cite{1511.00853}.

Furthermore, though the classification of black holes based on the range of charge-to-mass ratio is model-dependent, we can estimate a wide limit of the parameter $\alpha$ (or $\beta$) for a stable configuration with a timelike boundary. From eqs.~(\ref{rho}), (\ref{M}), and (\ref{Q}), we obtain
\begin{align}
\int_0^{r_s} A_{n-2}r^{n-2}\rho(r_s)dr&=\frac{2\pi Q^2}{(n-1)A_{n-2}r_s^{n-3}},\\
\int_0^{r_s} A_{n-2}r^{n-2}\rho(r)dr&=M-\frac{2\pi Q^2}{(n-3)A_{n-2}r_s^{n-3}},
\end{align}
which leads to
\begin{align}
r_s^{n-3}\geq\frac{4\pi(n-2)}{(n-1)(n-3)A_{n-2}}\frac{Q^2}{M}
\label{r_s_l}
\end{align}
when the weak energy condition is considered, i.e., $\rho>0$ and $\rho'(r)<0$, see subsection 2.1 for the details of analyses.
As a result, in order to keep the charged sphere inside the inner horizon ($r_s<r_-$), eq.~(\ref{r_s_l}), together with eq.~(\ref{r_s^<r_-^}), provides the range of the reduced parameter $\beta$,
\begin{align}
\frac{1}{4\pi}\frac{(n-1)(n-3)^2}{(n-2)^3}<\beta \leq\frac{1}{4\pi}\frac{(n-3)}{(n-2)}.
\end{align}

This inequality manifests that $\beta$ is lower bounded, which agrees with \cite{1706.03454} the 4-dimensional case, and more importantly that the range of $\beta$ grows narrow as the dimension becomes large.

\vskip 4mm

{\em Note: This work was begun when the first author was graduate student at Nankai University.}

\section*{Acknowledgments}
Y.-MW would like to thank L. Zhao for helpful discussions, and she acknowledges the financial support from the Research Foundation for Middle-aged and Young Teachers of Fujian Province under grant No.JT180747.
Y.-GM would like to thank H.P. Nilles for the warm hospitality during his stay at University of Bonn, and he acknowledges the financial support from the Alexander von Humboldt Foundation under a renewed research program and from the National Natural Science Foundation of China under grant Nos. 11675081 and 12175108.

\end{document}